\newif\ifMapleTrans
\MapleTransfalse
\ifMapleTrans
\documentclass[acmsmall,screen]{MapleTrans}
\usepackage[normalem]{ulem}

\usepackage{enumitem}
\usepackage{etoolbox}
\usepackage{breqn}
\usepackage{lineno}
\usepackage{maple}
\usepackage{xfrac}
\usepackage{xcolor}
\definecolor{mygreen}{rgb}{0,0.6,0}
\definecolor{mygray}{rgb}{0.5,0.5,0.5}
\definecolor{mymauve}{rgb}{0.58,0,0.82}
\definecolor{altblue}{rgb}{0.0,0.6,1.0}
\definecolor{lstbg}{cmyk}{0.07, 0.04, 0.18, 0}
\definecolor{morebluish}{cmyk}{0.06,0.04,0,0}
\usepackage{listings}
\input{listings-maple-definition.sty}
\setcopyright{acmlicensed}
\copyrightyear{2024}
\acmYear{2024}
\acmDOI{10.5206/mt.vViI.XXXXX}
\acmJournal{MAPLETRANS}
\acmVolume{X}
\acmNumber{Y}
\acmArticle{Z}
\acmMonth{0}
\else
\documentclass[11pt]{article}
\usepackage{fullpage}
\usepackage{amsmath,amssymb,amsfonts,amsthm}
\newtheorem{theorem}{Theorem}
\fi

\usepackage{cite}  
\usepackage{array} 
\usepackage{comment}
\newcommand{\mtwo}[4]{\ensuremath{\left [ \,\begin{array}{@{}c@{\;\,}c@{}}#1 & #2 \\ #3 & #4\end{array}\,\right ]}}
\newcommand{\bigmtwo}[4]{\ensuremath{\left [ \,\begin{array}{@{}cc@{}}#1 & #2 \\[1ex] #3 & #4\end{array}\,\right ]}}
\newcommand{\inv}[1]{\ensuremath{#1^{-1}}}
\title{Algorithms for Recursive Block Matrices}
\ifMapleTrans
\author{Stephen M. Watt}
\affiliation{%
\institution{University of Waterloo}%
\city{Waterloo Ontario}%
\country{Canada}%
\address{~}
\postcode{N2L 3G1}
}
\email{smwatt@uwaterloo.ca}
\else
\author{%
  Stephen M. Watt\\
  University of Waterloo\\
  \texttt{smwatt@uwaterloo.ca}
}
\date{May 20, 2024}
\fi
\begin{document}
\ifMapleTrans\else\maketitle\fi
\begin{abstract}
    We study certain linear algebra algorithms for recursive block matrices. 
    This representation has useful practical and theoretical properties.
    We summarize some previous results for block matrix inversion and present some results on triangular decomposition of block matrices.  
    The case of inverting matrices over a ring that is neither formally real nor formally complex was inspired by Gonzalez-Vega \textit{et al.}
\end{abstract}
\section{Introduction}
Algorithms on block matrices have both useful theoretical and practical properties.
From a theoretical point of view, algorithms for matrices with non-commuting elements allow recursive formulation of linear algebra problems, simplifying complexity analysis.  
Strassen's seminal result on matrix multiplication~\cite{strassen-mult} is perfect example.
From a practical point of view, block matrices provide a middle ground that avoids pathological communication bottlenecks in row-major or column-major code~\cite{dongara-scalability}.   
Recursive block matrices allow both dense and structured matrices to be represented with reasonable efficiency~\cite{gargantini-quadtree,abdali-wise-quadtrees}.
For these reasons, having recursive block matrix representations in mathematical software is desirable.

Modern programming languages used in mathematical computing, including \texttt{C++}, Julia, Python and Fortran 2023, provide data abstraction mechanisms that can support recursive block matrices in a natural way.  However, in designing a library for one such language, Aldor~\cite{watt-aldor}, that the standard algorithms for block matrices sometimes require breaking the block abstraction.
In building a type of block $n\times n$ matrices over a ring $R$, providing  ring operations on $R^{n \times n}$ is straightforward.
However providing a partial function for matrix inverse is not.  
The usual formulation to invert a $2\times 2$ block matrix requires at least one block and its Schur complement to be invertible.  
But this may not be the case --- a nonsingular matrix may consist entirely of singular blocks.   In situations such as this, the standard methods break the block abstraction and work on rows, \textit{e.g.}~\cite{aho-hopcroft-ullman}.

In this article we explore algorithms for recursive block matrices that respect the block abstraction.  We begin by summarizing earlier results for a matrix inversion method. The method is directly applicable to matrices over a ring with a formally real sub-field.  The method may be generalized to matrices over other rings using a technique suggested by Gonzalez-Vega \textit{et al.}~\cite{gema-lalo-henri}.  Following this, we undertake some exploration into triangular decomposition of recursive block matrices.

\section{Inversion of Recursive Block Matrices}
\label{sec:block-inverse}
Most ring operations on block matrices may be performed in a straightforward manner using only block operations.  
That is, for a block matrix
\[
M= \mtwo ABCD \in R^{2n\times 2n}
\]
only ring operations on $A, B, C, D \in R^{n\times n}$ are needed.
If all of the blocks of $M$ are invertible, the inverse of $M$ may be computed as
\[
M^{-1} = \bigmtwo
   {(A-BD^{-1}C)^{-1}}
   {(C-DB^{-1}A)^{-1}}
   {(B-AC^{-1}D)^{-1}}
   {(D-CA^{-1}B)^{-1}}.
\]
In practice, only two inverses are required---that of $A$ and its Schur complement, $S_A = D-CA^{-1}B$,
\begin{align}
M^{-1}
&=          \mtwo{I}{-A^{-1}B}{0}{I}
            \mtwo{A^{-1}}{0}{0}{S_A^{-1}}
            \mtwo{I}{0}{-CA^{-1}}{I}
= \bigmtwo
            {A^{-1}+A^{-1}BS_A^{-1}CA^{-1}}
            {-A^{-1}BS_A^{-1}}
            {-S_A^{-1}CA^{-1}}
            {S_A^{-1}}.
            \label{eqn:block-inverse}
\end{align}
If $A$ is not invertible, then a similar formula involving the inverse of another block and its Schur complement may be used, perhaps after a permutation of rows or columns.
The problem with this approach is that $M$ may be invertible even when all of $A$, $B$, $C$ and $D$ are singular.  In this situation, permuting the blocks is of no help. 
One approach is to break the block abstraction and use operations on whole rows of $M$ viewed as a flat $2^{2n\times 2n}$ matrix~\cite{aho-hopcroft-ullman}.

In earlier work~\cite{watt-block-inv}, we have shown a recursive algorithm to invert  matrices respecting a block abstraction.
In particular, row operations are not required for pivoting or otherwise.
The technique is to use the Moore-Penrose inverse so that the principal minors are guaranteed to be invertible and equation~\eqref{eqn:block-inverse} may be used. 
We summarize those results here.

We use the notation $R^{[2^k\times 2^k]}$ to mean the ring of $2^k \times 2^k$ matrices with elements in $R$, structured in recursive $2\times2$ blocks.  Any $n\times n$ matrix may be easily be embedded in such a ring.

\begin{theorem}
\label{thm:real}
If $R$ is a formally real division ring and $M \in R^{n\times n}$ 
is invertible, then it is possible to compute $M^{-1}$ as $(M^TM)^{-1}M^T$
using only block operations.
\end{theorem}
\noindent
By {\sl block operations}, we mean ring operations in $R^{[2^{k-1}\times 2^{k-1}]}$.
Examples of formally real rings are $\mathbb Q$, $\mathbb R$, $\mathbb Q[\sqrt 2]$ and $R[x,\partial]$ for formally real $R$.
\begin{theorem}
Let $C$ be a division ring with a formally real sub-ring $R$
and involution ``$*$'', such that for all $c \in C$, $c^* \times c$ is 
a sum of squares in $R$. 
If $M \in C^{[2^k\times 2^k]}$ is invertible, then it is possible to 
compute $M^{-1}$ as $(M^* M)^{-1} M^*$
using only block operations.
\end{theorem}
\noindent
Examples of such rings are the complexification of a formally real ring $R$ as $R[i]/\langle i^2 +1\rangle$ or quaternions over $R$ with the involution $(a + bi + cj + dk)^* = a - bi - cj -dk$.

The next result follows an observation of Laureano Gonzalez-Vega, using a technique of~\cite{gema-lalo-henri,mulmuley}.
\begin{theorem}
Let $K$ be a field.  If $M \in K^{2^k\times 2^k}$ is invertible, 
then it is possible to compute $M^{-1}$ as $(M^\circ M)^{-1} M^\circ$
using only block operations.
\end{theorem}
\noindent
Here {\sl block operations} mean ring operations in $K(t)^{[2^{k-1}\times 2^{k-1}]}$ and
$M^\circ = Q_n^{-1} M^T Q_n$ is a group conjugate of $M^T$,
with $Q_n = \mathrm{diag}(1, t, \ldots, t^{n-1})$.

In all cases, the time complexity is that of two $2^k\times 2^k$ matrix multiplications and one inversion using \eqref{eqn:block-inverse}.  
The inversion may be achieved with two $2^{k-1}\times2^{k-1}$ inversions and two $2^{k-1}\times 2^{k-1}$ multiplications to compute $S_A$ and its inverse, and four $2^{k-1}\times 2^{k-1}$ multiplications, namely
\begin{align*}
t_1 &= C\cdot A^{-1}  &
t_2 &= A^{-1} \cdot B &
t_3 &= t_2 \cdot S_A^{-1} &
t_4 &= t_3 \cdot t_1,
\end{align*}
relying on the symmetries $(M^T M)_{ji} = (M^T M)_{ij}$, $(M^* M)_{ji} = (M^* M)_{ij}{}^*$ and 
$(M^\circ M)_{ji} = t^{i-j} (M^\circ M)_{ij}$ for $j \ge i$.
Thus,
\[
T_{\mathrm{inv}}(2^k) = 2 T_\times(2^k) + 2T_{\mathrm{inv}}(2^{k-1}) + 4 T_\times(2^{k-1})
\]
where $T_\times(n)$ is the time complexity to multiply two $n\times n$ matrices.  If $T_\times(n) = \alpha n^\omega$ and $T(1)=1$, then
\[
T_{\mathrm{inv}}(n) = 2 \alpha n^\omega - (2\alpha - 1) n 
+ \frac{8\alpha (2^\omega +2) (n^\omega - n)}{4^\omega -4} 
\in O(T_\times(n)).
\]
This complexity was not spelled out in~\cite{watt-block-inv}.

\section{LU Decomposition of Recursive Block Matrices}
It is often desirable to factor \(M\) as \(L \cdot U\) where \(L\) and \(U\) are respectively lower- and upper-triangular.

Let \(M\) be a nonsingular $2\times2$ square block matrix
\[ M = \mtwo ABCD.\]  We show how to compute such lower-triangular $L$ and upper triangular $U$ such that $M = PLUQ$ with permutation matrices $P$ and $Q$.

\subsection{Assuming a Nonsingular Block}
If one of \(A\), \(B\), \(C\) or \(D\) is nonsingular,
permute the rows and columns of \(M\) as necessary to obtain \(M^\prime\) with nonsingular \(M^\prime_{11}\).

With \(T = \mtwo 0II0\), let \(M^\prime = P_1 \, M \, Q_1\)
where
\begin{align*}
    P_1 &= \begin{cases} I & \text{when \(A\) or \(B\) nonsingular}\\
    T &\text{otherwise}
    \end{cases} & 
    Q_1 &= \begin{cases} I & \text{when \(A\) or \(C\) nonsingular} \\
    T &\text{otherwise.}
    \end{cases}
\end{align*}
If, at the recursive step, $A^\prime$ is arranged to $A^{\prime\prime}$ so $A^{\prime\prime}_{11}$ is nonsingular, then we will have $M^{\prime\prime} = P_2 P_1 M Q_1 Q_2$, \textit{etc.}
We can now drop the \(P_i\), \(Q_i\), and the primes in what follows.
\subsection{Block LU Decomposition}
It is well-known that \(M\) may be factored in an LDU decomposition as
\begin{equation}
    \bigmtwo ABCD = \bigmtwo I0{CA^{-1}}I \bigmtwo A00{S_A} \bigmtwo I{A^{-1}B}0I.
    \label{eqn:block-lu}
\end{equation}
It is then possible to multiply the block diagonal middle factor on either the left or right, depending on desired convention, to obtain a block LU decomposition.   

We are interested, however, in an LU decomposition of \(M\). Regardless of whether we multiply the middle factor to the left or right, the block LU decomposition will in general have non-triangular \(A\) as the $(1,1)$ component of either \(L\) or \(U\).  
So \eqref{eqn:block-lu} is not what we want.   

We could iterate the process and next find LDU decompositions of \(A\) and  and \(S_A\), and so on, combining lower triangular matrices on the left and upper triangular matrices on the right. In this situation some of the intermediate multiplications will be specialized, and it is more convenient to consider LU decomposition directly.

\subsection{LU Decomposition}
The problem is to find lower and upper triangular matrices \(L\) and \(U\)
such that
\begin{align*}
    M &= LU &
    L &= \mtwo{L_1}0X{L_2}  & U &= \mtwo {U_1}Y0{U_2}
\end{align*}
with \(L_1\) and \(L_2\) lower triangular and \(U_1\) and \(U_2\) upper triangular.   This LU decomposition is different from the \(LDU\) decomposition which has \(L\) and \(U\) as above, but with \(L_i = U_i = I\) and \(D\) diagonal.  It is therefore suitable for recursive application.

The four components of \( M = L U\) are
\begin{align*}
A &= L_1 U_1 & B &= L_1 Y \\
C &= X\; U_1 & D &= L_2 U_2 + XY.
\end{align*}
Assuming \(A\) is nonsingular, these give two recursive LU decompositions
\begin{align*}
    A &= L_1 U_1 & L_2 U_2 &= D-XY\\
\end{align*}
in which case \(X\) and \(Y\) may be obtained as
\begin{align*}
    X &= C \inv{U_1} & Y = \inv {L_1} B.
\end{align*}
Note that if \(A\) is nonsingular, so are \(L_1\) and \(U_1\).

The matrices \(L_i\), \(U_i\), \(X\) and \(Y\) may be computed as:
\begin{align*}
L_1 U_1 & = A \\
X &= C U_1^{-1} \\
Y &= L_1^{-1} B \\
L_2 U_2 &= D-XY.
\end{align*}
The degree of freedom in choosing the diagonal elements of \(L\) and \(U\) is handled by taking the same convention in the recursive computations of \(L_1 U_1\) and \(L_2 U_2\).

\subsection{Complexity Analysis}
We have shown, subject to nonsingularity of a block, that the LU decomposition of an \(n\times n\) matrix may be computed with \(T_{LU}(n)\) multiplications using:
\begin{itemize}
    \item 2 size $n/2$ LU decompositions, \(2 T_{LU}(n/2)\),
    \item 2 size $n/2$ triangular matrix inversions, \(2T_{\inv \triangle}(n/2)\).
    \item 1 size $n/2$ general matrix multiplication, \(T_{\times}(n/2)\).
    \item 2 size $n/2$ triangular times general matrix multiplications, \(2T_{\triangle\times}(n/2)\).
\end{itemize}
The number of multiplications for LU decomposition by this method is therefore
\[
T_{LU}(n) = 2T_{LU}(n/2) + 2T_{\inv \triangle}(n/2) + T_\times(n/2) + 2T_{\triangle\times}(n/2).
\]

The number of multiplications and divisions to invert a triangular matrix is
\(
    T_{\inv \triangle}(n) = \frac 12 n(n+1).
\)

To multiply a triangular and general matrix of size \(n\times n\), form
\[
\mtwo ABCD \mtwo U0VW = \mtwo{AU+BV}{BW}{CU+DV}{DW}
\]
where \(U\) and \(W\) are lower triangular.
So
\(
T_{\Delta\times}(n) = 4 T_{\Delta\times}(n/2) + 2 T_\times(n/2).
\)
If \(T_\times(n) = \alpha n^\omega\), then
\[
T_{\triangle\times}(n) = \frac{2\alpha }{2^\omega-4} (n^\omega -n^2) + n^2
\]
and the LU decomposition method requires a number of multiplications 
\begin{align}
T_{LU}(n) &= 
\alpha  (n^\omega-n) \frac{2^\omega}{(2^\omega-2)(2^\omega-4)}
+ (n^2-n) \left (
    \frac 32 -\frac{2\alpha }{2^\omega-4}
  \right)
+ \tfrac 12 n\log_2 (n)
+n \in O(T_\times(n)).
\label{eqn:lu-general}
\end{align}

\subsection{When All Blocks Are Singular}
It remains to handle when $A$, $B$, $C$ and $D$ are all singular.
If $M$ is nonsingular and its elements are from a sufficiently large domain, then one approach would be through randomization.   Letting $R_L$ and $R_U$ be random lower and upper triangular matrices, one can with high probability compute $L^\prime U^\prime = R_L M R_U$ as above, so $L = R_L^{-1} L^\prime$ and $U = U^\prime R_U^{-1}$.  
Also, note that LU decomposition is meaningful when $M$ is singular~\cite{jeffrey-lu}.  This remains a topic of on-going work.

\IfFileExists{IfExistsUseBBL.tex}{%

}{%
\bibliography{main.bib}
\bibliographystyle{plain}
}
\end{document}